\documentclass[12pt]{article}
\usepackage{pdproc,epsfig} 
\newcommand{\quq}{\theta_{13}}
\newcommand{\dmatmo}{\Delta m^2_{32}}

\newcommand{\nme}{\nu_\mu \rightarrow \nu_e}

\newcommand{\neb}{\bar{\nu}_e}
\newcommand{\sinquq}{\sin^2(2\theta_{13})}
\newcommand{\mxang}{\sin^2(2\theta)}
\begin{document}

\renewcommand{\thefootnote}{\alph{footnote}}
  
\title{ 
SEARCHING FOR THE NEUTRINO MIXING ANGLE $\quq$ AT REACTORS
\footnote{submitted to the proceedings of the XIIth International
Workshop on Neutrino Telescopes, Venice 2007}
}
\author{MAURY GOODMAN}
\address{ High Energy Physics Division
\\
Argonne National Laboratory
\\
Argonne IL 60439\\
 {\rm E-mail: maury.goodman@anl.gov}}
\abstract{
Two neutrino mixing angles have been measured, and much of the
neutrino community is turning its attention to the unmeasured
mixing angle, $\quq$, whose best limit comes from the reactor
neutrino experiment CHOOZ.\cite{bib:chooz}  New two detector
reactor neutrino experiments are being planned, along with 
more ambitious accelerator experiments, to measure or further
limit $\quq$.  Here I will overview how to measure $\quq$
using reactor neutrinos, mention some experiments that were
considered and are not going forward, and review the current
status of four projects: Double Chooz in France, Daya Bay
in China, RENO in South Korea and Angra in Brazil.  Finally I
will mention how the neutrino observer can gauge progress in
these projects two years from now as we approach the times
corresponding to early
estimates for new results.
}
\normalsize\baselineskip=15pt

\section{Introduction}
\par Since 2003, there has been a worldwide effort to plan new
reactor neutrino experiments using two or more detectors to measure
or further limit the only unmeasured neutrino mixing angle, $\quq$.
The best current limit on $\quq$ comes from the reactor experiment
CHOOZ,\cite{bib:chooz} which ran in the 1990's along with 
Palo Verde\cite{bib:palo} to determine if the atmospheric neutrino
anomaly could be explained with $\theta_{12}$.  Here I will describe some
features and the current status of reactor neutrino experiments,
which I expect to be the first to improve our knowledge of
$\quq$ further.  The reader whose
only interest is in the status of current
projects can skip to 
Section 3.2.

\par In a sense, reactor neutrino $\neb$ disappearance
experiments are complementary to the new off-axis accelerator $\nu_e$
appearance experiments, T2K \& NO$\nu$A\cite{bib:t2k,bib:nova}, whose
goal is also to study $\quq$.  The magnitude of a $\quq$ signal
at a new reactor
neutrino experiment is affected only by the
uncertainty of the value of $\dmatmo$, which is currently bounded by
$2.48 < \dmatmo < 3.18 \times 10^{-3}eV^2/c^4$.\cite{bib:minos,bib:trish}
On the other hand,
the ability of an accelerator experiment to measure $\quq$ is also
affected by the uncertainty in $\theta_{23}$, 
$ 0.36 < \sin^2(\theta_{23}) < 0.63$\cite{bib:pdg}, and 
the uncertainty in the CP violating
phase $\delta$, $0 < \delta < 2\pi$.  Thus a precise measurement
of $\quq$ by both reactor and accelerator experiments could be
used to constrain $\theta_{23}$ and/or $\delta$.  On the other hand,
a failure to find a value for $\quq$
by the reactor experiments will have a
rather negative effect on the expected physics capabilities for
the accelerator experiments which are much more expensive.  For example, a
limit of $\sin^2 (2 \quq) < 0.02$, which reactor experiments could
achieve before T2K or NO$\nu$A start running, would mean
 that the accelerator experiments, even with 
increases in beam power, could not measure evidence for matter effects
or CP violation and would only have a narrow window to 
find evidence for a non-zero $\quq$

\begin{figure}
\vspace*{3pt}
\begin{center}
         \mbox{\epsfig{figure=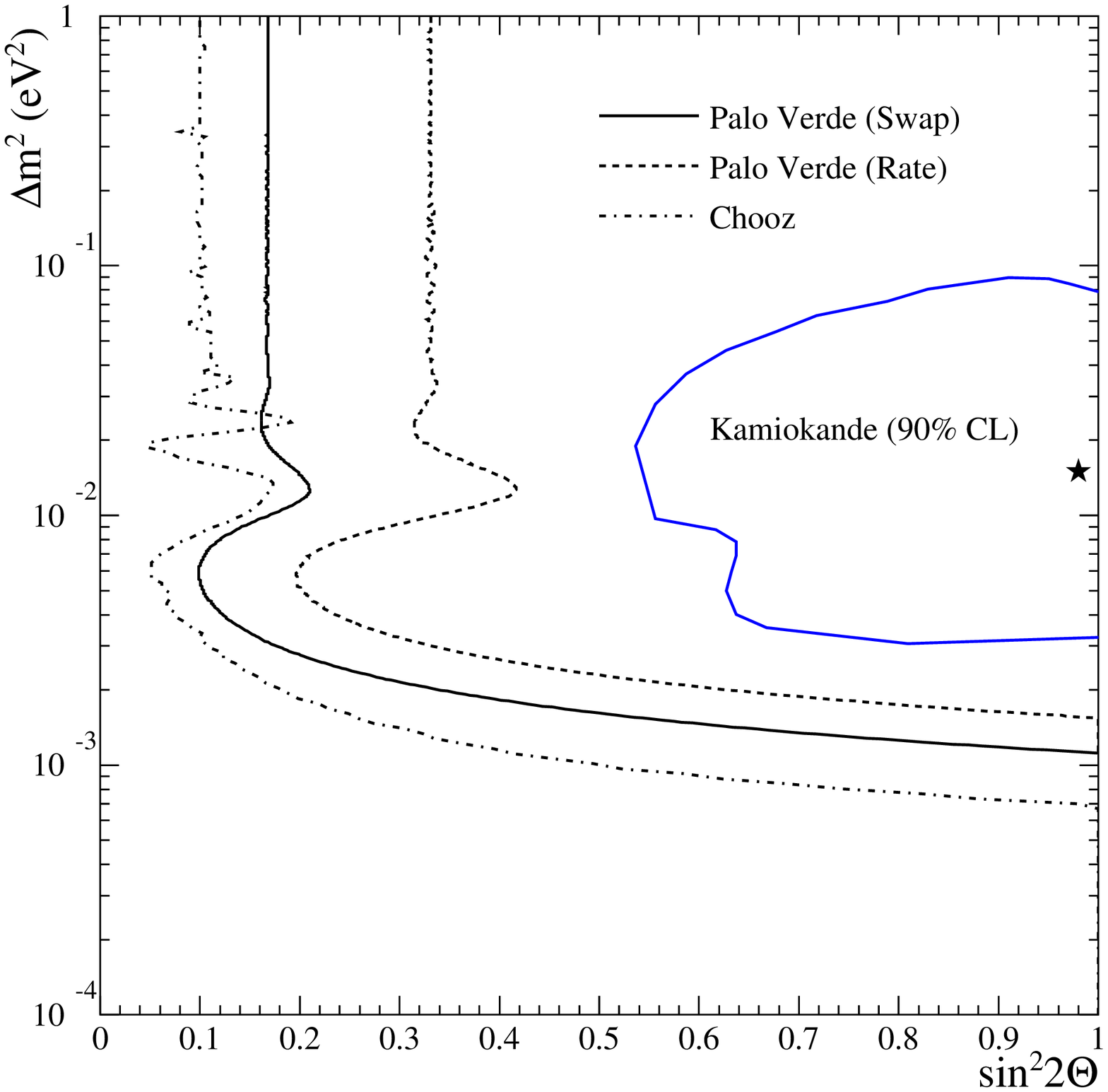,width=8.0cm}}
\caption{The CHOOZ and Palo Verde Limits on $\quq$ as a function of $\dmatmo$.}
\label{fig:chooz}
\end{center}
\end{figure}

The current limit on $\quq$ from the Chooz experiment is shown
in Figure \ref{fig:chooz}.  The analysis was actually done for
$\theta_{12}$, but given the value of $\Delta m^2_{21}$, it serves
as a valid analysis for $\quq$.  The curve shows the 90\% CL allowed
and prohibited values of $\sinquq$ as a function of $\dmatmo$.
In order to compare experiments, it 
is common to quote a single number as
the $\quq$ limit, but this requires a few assumptions.  
As a consequence, a large variety of numbers are
quoted as the CHOOZ limit, such as $\sinquq < 0.10, 0.11, 0.14, 0.15, 0.20$.
One cause for this is the time-dependence of the $\dmatmo$ measurement of
Super-K and now MINOS.  There is also no unique method of picking
the value of $\dmatmo$ to use.  (The union of two CL curves is not
a Confidence Level.)  While the ``best fit" $\dmatmo$ value is often chosen,
the PDG has elected to use the one sigma low value of $\dmatmo$ where
the larger value of $\quq$ is achieved, and they
obtain $\sinquq < 0.19$.   Another
more mundane issue which requires care is that, depending on the 
application, $\quq$ is often expressed in degrees, in radians, as
$\sin(\quq)$, $\sin^2(\quq)$, $\sin^2(\theta_{\mu e})$, 
$\sinquq$ and $U_{e3}$.  The relationships between these
expressions are simple, but factors of
two errors are common.  Finally, since comparison with the
sensitivity of future accelerator
experiments is often made, note that the accelerator 
experiments have additional ambiguities and degeneracies in interpreting
a $\quq$ limit from $\theta_{23}$, $\delta$ and the mass 
hierarchy.\cite{bib:lindneramb}

\section{The planning of a new generation of reactor neutrino
experiments}
The ingredients for the design of a
new reactor $\quq$ experiment have been laid out in detail in the 
white paper, ``A new reactor neutrino experiment to measure $\quq$"
prepared by an International Working Group comprised of 125 authors from
40 institutions in 9 countries\cite{bib:white}.  The optimum location
for the far detector depends on $\dmatmo$, and also on the
experiment's ultimate exposure.  Sites from 1.1 to 2.0 km have been
chosen.  The near detector needs to be located
close to the core to measure the unoscillated spectrum.  Local factors,
such as reactor access and topological features modify where 
detectors will be placed.  Important general features of reactor
experiments are the effects of luminosity on the sensitivity,
detector design, scintillator stability, calibration, backgrounds
and systematic errors.

\par
\begin{figure}
\vspace*{3pt}
\begin{center}
         \mbox{\epsfig{figure=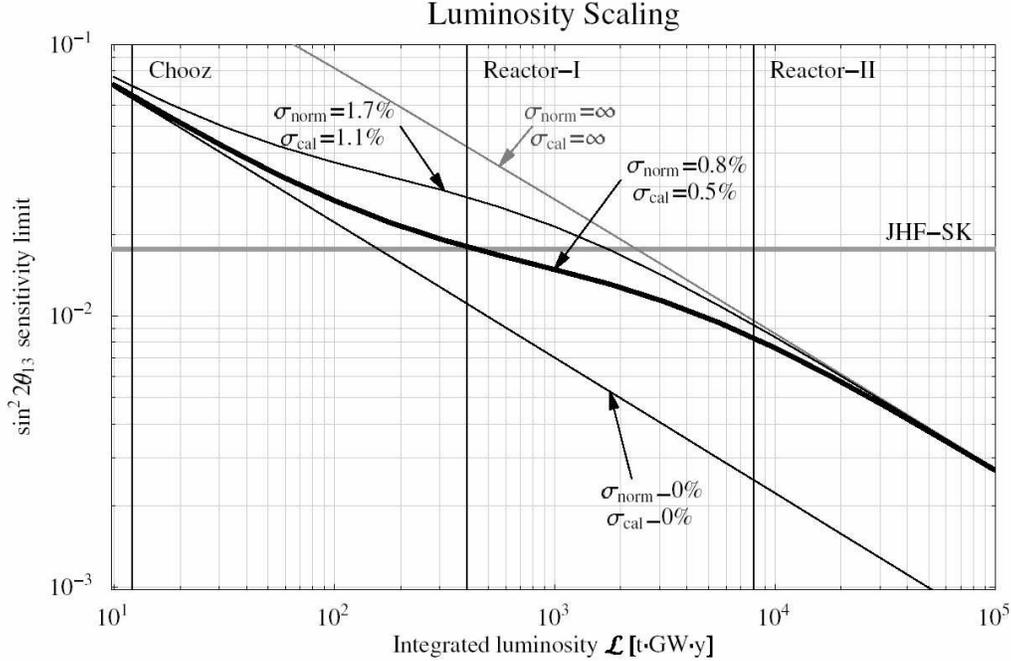,width=14.0cm}}
\caption{Luminosity scaling for a reactor experiment's sensitivity
for $\quq$.  The solid curve
shows what might be achieved with reasonable assumptions for systematic
error.}
\label{fig:lindner}
\end{center}
\end{figure}

 The neutrino oscillation sensitivity for a reactor neutrino experiment
comes from measuring a smaller number of neutrinos than would be expected
if $\quq=0$, and measuring an energy distribution consistent with 
$\neb$ disappearance due to oscillations.  These can be called the ``rate" test and the
``shape" test, but every experiment will use all available
information.
The effective ``luminosity" for a reactor 
experiment can be expressed in GW-ton-years, or the product of the
reactor's thermal power times the size of the detector times the length
of time the detectors operate.  An example of how the sensitivity of an
experiment varies with luminosity is given in Figure \ref{fig:lindner}.
Two extreme examples, which represent straight lines on this log-log
plot, are for no systematic error, and for infinite systematic error
in normalization and energy calibration.  In the latter case, an oscillation
signature is recognized by the appropriate wiggles in an energy distribution.
Such a signal would be affected by bin-to-bin systematic errors, but
not by the same systematic errors which limit the ``rate" test.  
Two other curves are drawn with possibly realistic estimates of
systematic error for the next round of experiments.
Vertical lines are drawn at 12 GW-ton-years, corresponding to CHOOZ,
400 GW-ton-years, which could quickly and dramatically increase the
world's sensitivity, and a more ambitious project with 8000 GW-ton-years.

\par
\begin{figure}
\vspace*{3pt}
\begin{center}
         \mbox{\epsfig{figure=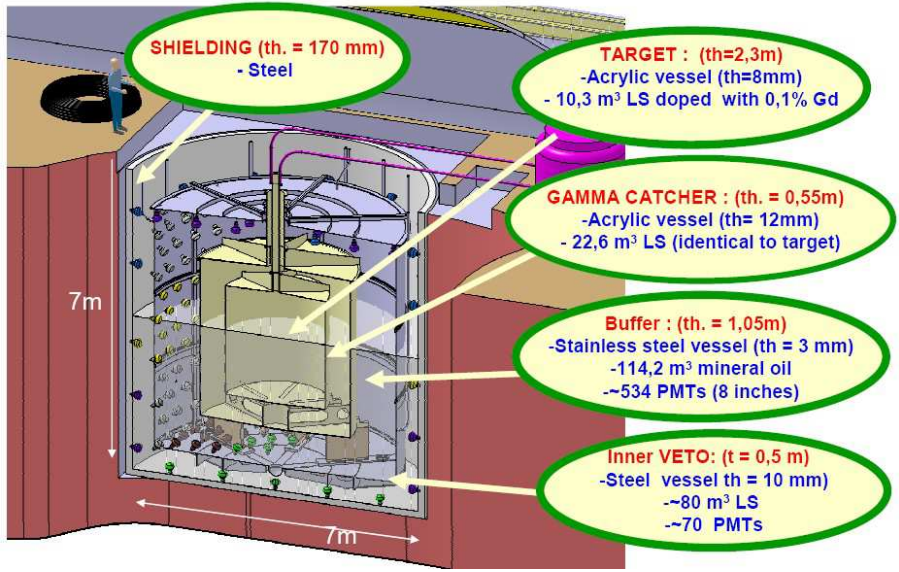,width=12.0cm}}
\caption{Plan for the Double Chooz detector(s) with three
optically connected volumes:  a target with 0.1\% Gd,
a $\gamma$-catcher with scintillator and no Gd, and a buffer
with mineral oil and no scintillator.}
\label{fig:choozdet}
\end{center}
\end{figure}

 CHOOZ had
a volume of Gd-loaded liquid scintillator, optically connected
and surrounded by 
liquid scintillator without Gadolinium.  New detector designs
involve the addition of a third volume of mineral oil without
scintillator, as shown in the Double Chooz design of 
Figure~\ref{fig:choozdet}.  An inner volume of Gd loaded scintillator
serves as a well-defined fiducial volume for neutrino interactions,
with a very high neutron capture cross section.  A second layer
of scintillator, called the ``$\gamma$-catcher", measures the energy
of any photons from positron annihilation or neutron capture which
escape the fiducial volume, and a third volume, or ``buffer", shields
the active volume from backgrounds originating in the rock or
phototubes.

\par The scintillator in CHOOZ showed a degradation of its
transparency over time, which resulted in a decrease of the light
yield.  Such a degradation would be unacceptable in a new experiment,
particularly if it differed between two detectors.  Suspicions concentrate
on possible contaminants which may have caused the Gd to come out of solution,
so that clean and robust liquid handling systems will be required to 
maintain good optical qualities.  Newly developed
scintillator formulations from groups at Heidelberg\cite{bib:dc}
 and Brookhaven\cite{bib:db}
have shown that it is possible
to satisfy the stability requirements for the long time periods needed.
\par Precise calibration will be necessary to ensure that
the response of two or more 
detectors is identical. This will be accomplished
by the introduction of radioactive sources that emit
gammas, electrons, positrons and neutrons.  Light flasher systems and
lasers will be used to test the stability of photo-detectors.  Cosmic ray
muons will also be used, and particular cosmogenic nuclei, such as 
$^12$B also can be used to provide a calibration.  An important
consideration is that identical calibration systems be used for all
detectors.  One possibility that has been proposed is to use
multiple and movable detectors, in order to increase the available
information regarding cross-calibration.
\par The neutrino signature is a coincidence between a prompt positron
annihilation and a delayed neutron capture with a mean life of
30 $\mu$s.
There are two kinds of backgrounds:
accidental ones where the two signals have different causes,
and correlated backgrounds.  Two important correlated 
backgrounds are fast neutrons, which can cause two
signals separated by a typical neutron capture time, and 
$^9$Li, which can be created by spallation when a muon passes
through the scintillator.  The danger of $^9$Li is that it has a
long decay time ($\sim$ 130 ms), and the decay
leads to both a neutron and an electron, creating a signal much
like a reactor neutrino.  The long decay time makes it unrealistic
to veto every throughgoing muon which might have been the cosmogenic
source.  While $^9$Li production has been measured,
its dependence on the muon energy is poorly known, so predictions
of the rates at a particular depth may not be accurate.
All correlated backgrounds can be reduced by putting the detectors
deep enough underground so that there are large overburdens, though this has a cost
in civil construction.
\par Finally, it is necessary to reduce the systematic errors for
counting reactor neutrino events below those that were achieved
by CHOOZ and Palo Verde.  The use of a second detector and the 
definition of the fiducial volume at the target/$\gamma$-catcher 
interface provide large reductions in systematic errors, and the
experiments have to be careful that other effects do not limit them
for their planned-for statistics.  A comparison of the
systematic error goals for Double Chooz and Daya Bay has been
tabulated by Mention et al.\cite{bib:mention} and is presented
in Table \ref{tab:syst}.  
\begin{table}[ht]
\caption{Comparison of Systematic errors 
for the CHOOZ analysis and estimates of the relative and
absolute errors in Double Chooz and Daya Bay, as tabulated by
Mention et al. in Reference 19.}
\vspace*{5pt}
\begin{tabular}{l|c|cc|ccc}\hline
Error Description & CHOOZ & ~~Double & Chooz~~~ & & Daya Bay & \\
& & & & & No R\&D & R\&D \\
& {\small Absolute} & {\small Absolute} & {\small Relative} & {\small Absolute} & {\small Relative} & {\small Relative}\\\hline \hline
Reactor & & & & & & \\ \hline
Production $\sigma$ & 1.90~\% & 1.90~\% & & 1.90~\% & & \\
Core powers & 0.70~\% & 2.00~\% & & 2.00~\% & & \\
Energy/fission & 0.60~\% & 0.50~\% & & 0.50~\% & & \\
Solid angle & & & 0.07~\% & & 0.08~\% & 0.08~\% \\ \hline
Detector & & & & & & \\ \hline
Detection $\sigma$ & 0.30~\% & 0.10~\% & & 0.10~\% & & \\
Target Mass & 0.30~\% & 0.20~\% & 0.20~\% & 0.20~\% & 0.20~\% & 0.02~\% \\
Fiducial volume & 0.20~\% &  & & & & \\
Target \% H & 0.80~\% & 0.50~\% & & ? & 0.20~\% & 0.10~\% \\
Dead time & 0.25~\% & & & & & \\ \hline
Analysis & & & & & & \\ \hline
$e^+$ escape & 0.10~\%  & & & & & \\
$e^+$ identification & 0.80~\% & 0.10~\% &  0.10~\% & & & \\
n escape & 0.10~\% & & & & & \\
n capture \% Gd & 0.85~\% & 0.30~\% & 0.30~\% & 0.10~\% & 0.10~\% & 0.10~\% \\
n identification & 0.40~\% & 0.20~\% & 0.20~\% & 0.20~\% & 0.20~\% & 0.10~\% \\
$\neb$ time cut & 0.40~\% & 0.10~\% & 0.10~\% & 0.10~\% & 0.10~\% & 0.03~\% \\
$\neb$ distance cut & 0.30~\% & & & & & \\
n multiplicity & 0.50~\% & & & & 0.05~\% & 0.05~\% \\ \hline \hline
{\bf Total} & {\bf 2.72~\%} & 2.88~\% & {\bf 0.44~\%} & 2.82~\% & 0.39~\% &
 0.20~\% \\ \hline\hline
\end{tabular}
\label{tab:syst}
\end{table}

\section{Sites for reactor experiments}
\subsection{Projects that were previously considered}
\par Four reactor $\nu$ projects have received some funding and
are moving forward.  These four experiments, which will be described
in the following sections, are Double Chooz, Daya Bay, RENO and Angra.
As members of the International Working Group considered locations
for new reactor experiments, a number of other
possible locations were studied
which have since been dropped.  It is instructive to consider some of
the strengths and weaknesses of sites that are not currently being
pursued.

\par The first idea for a two-detector experiment to measure $\quq$
was KR2DET\cite{bib:kr2det}.  This would have been built at the
Krasnoyarsk reactor in Russia, which was originally built underground
for producing weapons-grade plutonium.  Two 46 ton detectors would 
have been 115 m and 1000 m from the reactor.
Since the whole complex is 
underground, the 600 meters of water equivalent (m.w.e.)
 overburden would shield against a high rate
of cosmogenic nuclei, such as $^9Li$, and the near and far detector
would have the same low backgrounds.  Unfortunately, local officials
were not cooperative at the prospect of an international collaboration
at their formerly secret soviet city.
\par The reactor complex 
in the United States with the highest power, and the site of
a former reactor neutrino experiment, is Palo Verde, in Arizona.  The previous
collaboration had a poor decommissioning experience and the reactor
company was not approached about a new project.  There was a collaboration which
did an extensive site study at the Diablo Canyon reactor on the
coast of California.  The hills there offered an opportunity for
considerable overburden for both the near and far detector.  However PG\&E,
the reactor power company, had recently gone through a politically
motivated bankruptcy, and they decided not to cooperate with
the collaboration after the initial studies.  The most complete proposal
in the United States was put forward by a large collaboration at the
Braidwood reactor, about an hour's drive from Chicago\cite{bib:braidwood}.
  Good cooperation
with the Exelon Corporation was obtained after efforts from the Directors of
Fermilab and Argonne.  In Illinois, the overburden would need to
be achieved
by a vertical shaft rather than horizontal tunnels.  Although the per-foot cost for a shaft
is higher than a tunnel, the shaft height to reach a given overburden
can be obtained with less digging than
for the length of a typical mountain
tunnel, and civil construction costs are comparable.  An experiment was
designed which could move two pairs of 65 ton detectors between two 180 m
shafts about 1 km apart outside the reactor's security fence.  After
consideration by the Neutrino Scientific Assessment 
Group (NuSAG)\cite{bib:nusag},
the DOE decided not to fund the Braidwood experiment, presumably because
it was more expensive than the alternative, which was support for U.S.
participation in the Daya Bay project.

\par In Japan, a collaboration formed to prepare an experiment called
KASKA at the Kashiwazaki-Kariwa complex south of 
Niigata\cite{bib:kaska}.  With seven 3.4 GW$_{th}$ nuclear power plants,
it is the world's most powerful reactor complex.  The plants are 
located in two small clusters, so two near detectors were planned.  The
absence of hard rock at the desired depth led to a design
in which the detectors were placed in deep narrow shafts.
However the economics of shafts limits their size and hence the size of the
detector that could be placed in them.
The collaboration developed an excellent relationship with the nuclear
power company and conducted extensive boring studies 
to plan the shafts for 4.5 ton detectors.  The Japanese funding
agencies, which also support the KamLAND and Super-Kamiokande
experiments, decided not to support this project.

\newpage
\subsection{Double Chooz in France}
\label{sec:dc}
Double Chooz will use the location of the CHOOZ
experiment as its site for its far detector.  By avoiding civil 
construction costs for the far site, Double Chooz will be less
expensive and will be
able to get started more quickly than the alternatives.  There will be
a near detector location 270 m from the middle of the two reactor cores,
with an overburden of about 90 m.w.e., which conservatively maintains
a similar signal to background as the far detector.  
Engineering for a near site has been provided by the French
Electricity Company, Ed.F. The 
final design will be completed during 2007 and the
 lab will be ready in 2009.

\begin{figure}
\vspace*{3pt}
\begin{center}
         \mbox{\epsfig{figure=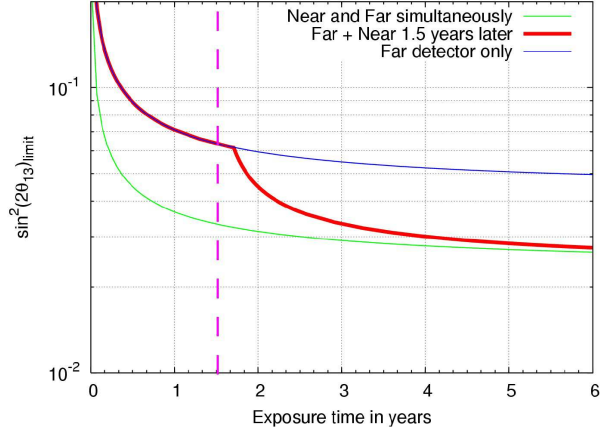,width=8.0cm}}
\caption{Expected Double Chooz $\quq$
sensitivity versus time.}
\label{fig:choozsched}
\end{center}
\end{figure}

The design for the three volume
Double Chooz detector was shown in 
Figure~\ref{fig:choozdet}.  A three volume prototype was built for
the R\&D stage, and the project is now entering
the construction stage.  Key parameters of the Double Chooz Experiment are
given in Table~\ref{tab:detsum}.  
Initial tenders
for the steel shielding and for the scintillator
have already been issued, and the far detector will be
installed and operated while the near detector lab is under
construction.  With just the 10 ton far detector, the CHOOZ limit on
$\quq$ can be passed in a few months.  When the near detector is
operational, the full sensitivity can be reached quickly, as shown
in Figure \ref{fig:choozsched}.

%

As the site of a former reactor neutrino experiment with extensive
reactor off running, Chooz is one place where backgrounds have been
measured.  Accidental backgrounds in Double Chooz will be much
lower than CHOOZ because sand will be replaced by 170 mm steel
shielding, and because of the buffer.
At the far detector, where 60 neutrinos per day will be 
measured, accidental backgrounds will be about 2 per day, while
correlated backgrounds from fast neutrons will be an order of magnitude
smaller.  The estimate for $^9$Li is 1.4 per day, based on measurements
in Chooz.  The near detector, which should measure 1012 neutrino events
per day, will have accidental backgrounds of about 22 per day, and 1.3 
per day from fast
neutrons.  The estimate for $^9$Li is 9 per day.
While Double Chooz will be both the cheapest experiment and the first to
provide new  knowledge on $\quq$, its lower ultimate sensitivity has been
used by some funding agencies to deny it the resources that could have
provided that knowledge in a more timely way.  

\begin{table}[htbp]
\caption[Detectors] {Summary of the some parameters  of the
proposed Double Chooz experiment.}
\label{tab:detsum}
\begin{center}
\begin{tabular}{lrc}
\hline
Thermal power & 4.27 GW & each of 2 cores \\
Electric power & 1.5 GWe & each of 2 cores \\
$\neb$ target volume & 10.3 m$^3$ & Gd loaded LS (0.1\%) \\
$\gamma$-catcher thickness & 55 cm & Gd-free LS\\
Buffer thickness & 105~cm & nonscintillating \\
Total liquid volume & $\sim$237~m$^3$ & \\
Number of phototubes per detector & 534 8{\tt "} & 13\% coverage \\
Far detector distance  & 1050~m & averaged\\
Near detector distance & 280~m & averaged\\
Far detector overburden & 300 m.w.e. & hill topology\\
Near detector overburden & 70--80 m.w.e. & shaft\\
$\neb$ 5 years far detector events & 75,000 & with a 60.5\% efficiency\\
$\neb$ 5 near detector events & 789,000 & with a 43.7\% efficiency\\
Relative systematic error        & 0.6\% & \\
Effective bin-to-bin error       & 1\%   & background systematics \\
Running time with far detector only & 1--1.5 year & \\
Running time with far+near detector & 3 years & \\
$\mxang$ goal in 3 years & 0.02--0.03 & (90\% CL) \\
\hline
\end{tabular}
\end{center}
\end{table}

\subsection{Daya Bay in China}
\begin{figure}
\vspace*{3pt}
\begin{center}
         \mbox{\epsfig{figure=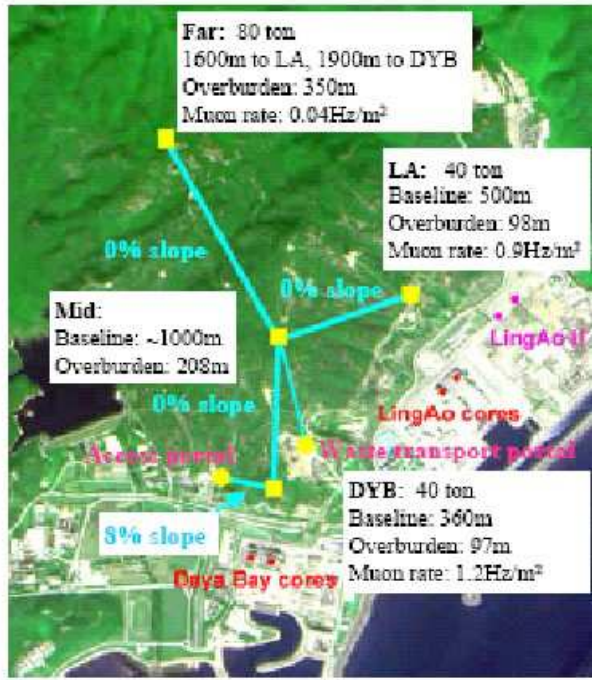,width=8.0cm}}
\caption{The layout of reactors and detectors at Daya Bay.}
\label{fig:daya1}
\end{center}
\end{figure}

The Daya Bay Complex, located near Hong Kong in Guangdong Province China,
currently consists of two pairs of reactors, called Daya Bay and Ling Ao.
The centers of each pair of reactors are about 1100 m apart.  In addition,
two more reactor cores near Ling Ao are under construction (Ling Ao II)
and should be in operation by 2011, resulting in a total 17.4 GW$_{th}$ reactor
power.  With this geometry, two near detectors are needed to monitor the reactor
power, as well as a far detector, as shown in Figure~\ref{fig:daya1}.
Important factors for the near sites are the estimated muon induced
backgrounds, which are a function of overburden.
The near sites were optimized using a global $\chi^2$, which takes into
account backgrounds, mountain profile, detector systematics and residual
reactor related systematics.  A summary of distances obtained
is provided in Table~\ref{tab:db}.
\begin{table}[ht]
\caption{Distances between reactors and planned detectors
at Daya Bay.}
\begin{center}
\begin{tabular}{|l||r|r|r|}\hline
~~~~~~Detectors & DB near & LA near & far \\
Reactors & (m) & (m) & (m) \\ \hline
DB cores & 363 & 1347 & 1985 \\ \hline
LA cores & 857 & 481 & 1618 \\ \hline
LA II cores & 1307 & 526 & 1613 \\ \hline
\end{tabular}
\label{tab:db}
\end{center}
\end{table}

The cylindrical Daya Bay detector will contain three zones, with a target,
$\gamma$-catcher and buffer, as described above.  The 224 phototubes
will be located on the sides of each 20 ton detector, 
with reflective surfaces at the top
and bottom.   The multiple detectors at each site will be used to
cross-calibrate each other, and the possibility of movable detectors is being
studied.  In Daya Bay's baseline design, the detectors at each site are
placed inside a large water buffer/water Cerenkov muon detector.
For the far hall, this is similar to a swimming pool with
dimensions 16 m  $\times$ 16 m $\times$ 10 m (high).  In addition,
water tanks of 1 m $\times$ 1 m are used as an outer muon tracker.  
\begin{figure}[ht]
\begin{minipage}[t]{0.60\textwidth}
\begin{center}
\includegraphics*[width=8cm,angle=0,clip]{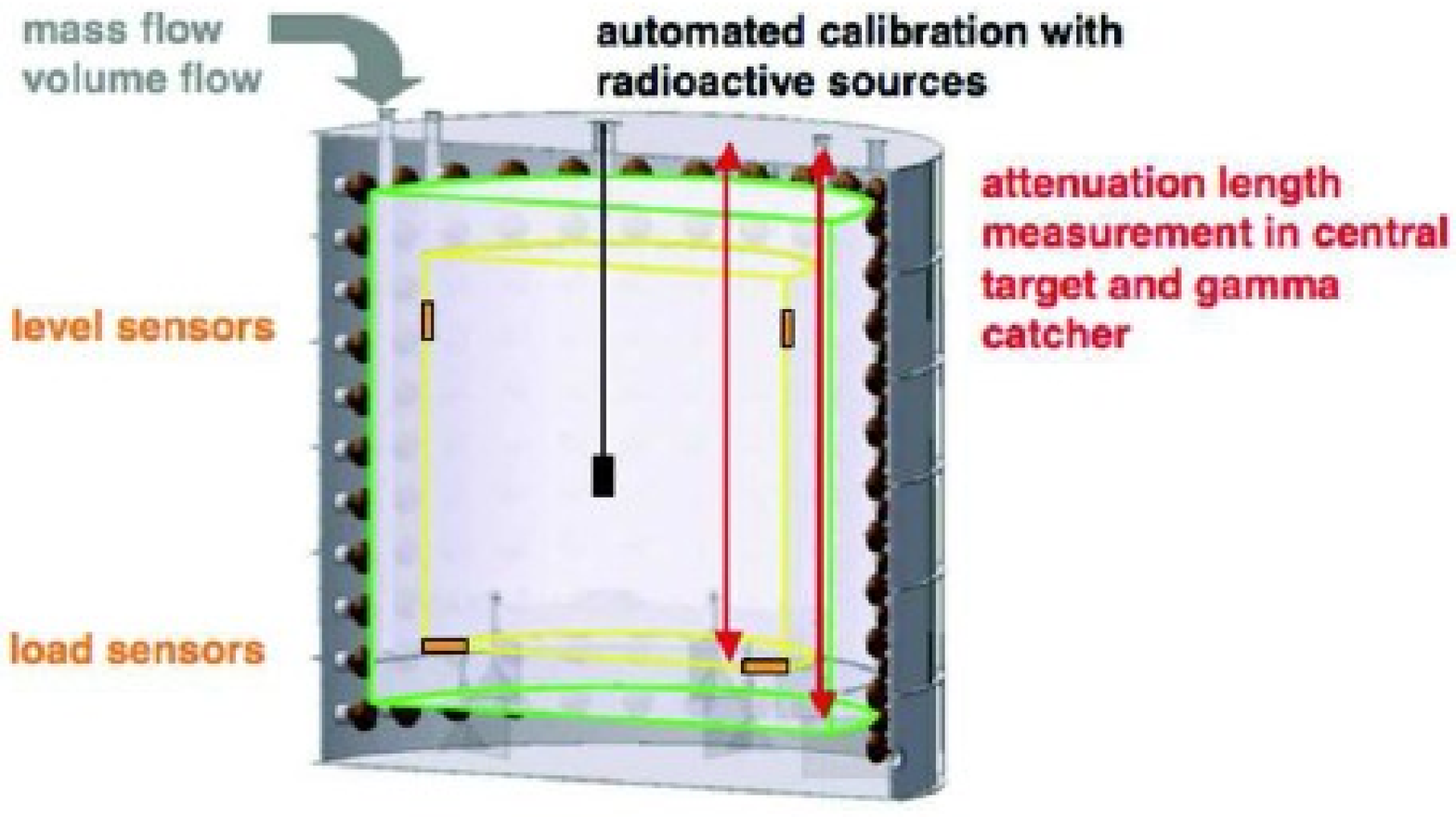}
\caption{\label {fig:daya2} 
Design of a Daya Bay Module showing a variety of
monitoring tools.}
\end{center}
\end{minipage}
\hfill
\begin{minipage}[t]{0.30\textwidth}
\begin{center}
\includegraphics*[width=4cm,angle=0,clip]{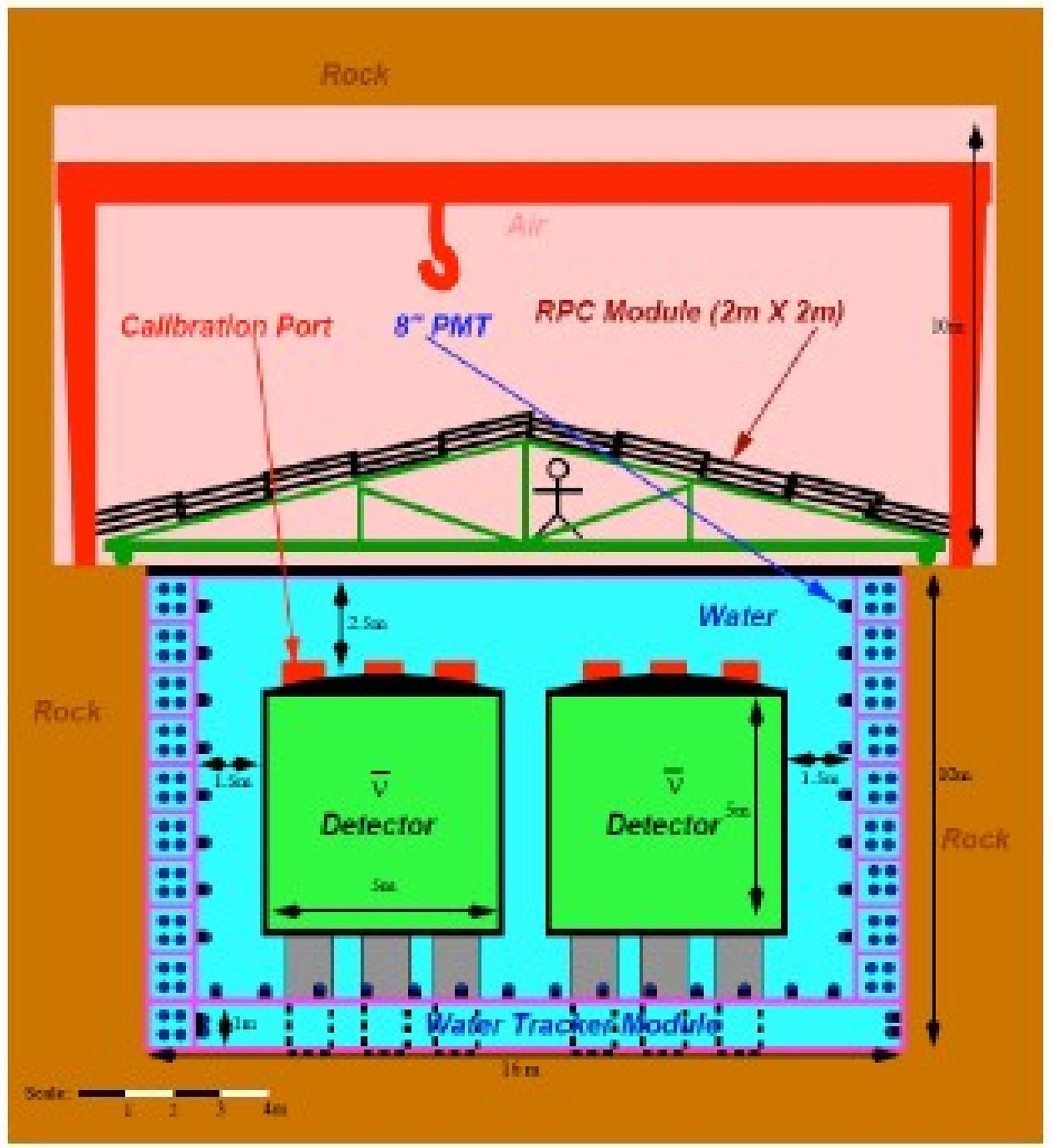}
\caption{\label {fig:daya3} 
Daya Bay Detectors in the water buffer/Veto System}
\end{center}
\end{minipage}
\hfill
\end{figure}
The large depth of the Daya Bay detectors will be used to
keep cosmogenic backgrounds at a small level.  Currently,
tests involving muon and neutron backgrounds are taking place
with a number of detectors at the Aberdeen tunnel in Hong Kong,
which has a similar overburden.
\par With the large reactor power and large overburden to reduce
backgrounds, Daya Bay is an excellent choice for a reactor $\quq$
experiment.  With support from the Chinese government and
the U.S. Department of Energy, it is poised to be an excellent
reactor experiment.  With a baseline detector systematic error of 0.38\%
and a goal of 0.18\%, they hope to take full advantage of the statistical
uncertainty of 0.2\%.  Data taking with two near halls and far hall
could begin in June 2010.  With three years of running, Daya
Bay will reach $\sinquq < 0.008$ or better.

\subsection{RENO in South Korea}
\begin{figure}
\begin{center}
         \mbox{\epsfig{figure=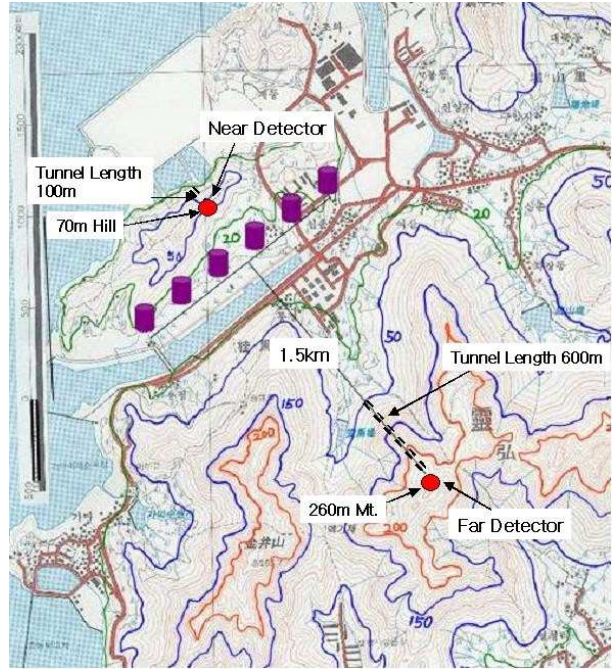,width=8.0cm}}
\caption{A topographic map of the YoungGwang site showing
the proposed locations of the near and far detectors for RENO.}
\label{fig:skreno}
\end{center}
\end{figure}

The South Korean Reactor Experiment for Neutrino Oscillation (RENO) collaboration 
is working on an experimental project at the YoungGwang reactor
complex, which consists of six equally spaced reactors in a line
on the west coast of South Korea.  
A schematic setup showing the topography and the proposed location
of the near and far detectors is shown in Figure \ref{fig:skreno}.
The near detector at a distance of
about 150 m would be under a 70 m high hill,
and the far detector at a distance of
1.5 km would be under a 260 m high mountain.
\par
\begin{figure}
\begin{center}
         \mbox{\epsfig{figure=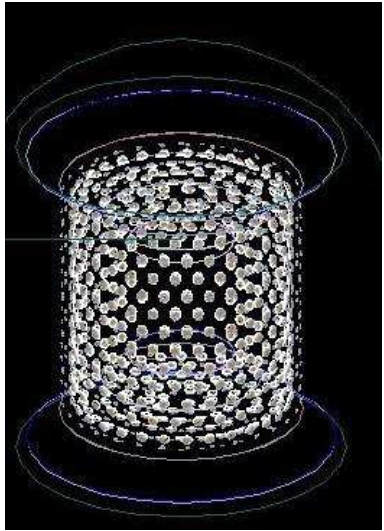,width=5.0cm}}
\caption{The phototube configuration in the RENO
GEANT simulation.}
\label{fig:reno}
\end{center}
\end{figure}

 After getting a \$9M funding approved by the government of
Korea, the RENO collaboration\cite{bib:reno} has been undertaking
detector design since May 2006.  Various samples of liquid scintillator
are under investigation with respect to the
 long-term stability of their optical
properties.  Other tests include compatibility with stainless steel
and mylar and an acrylic cracks test.  
A RENO prototype contained
50 liters of Gd loaded scintillator with a 400 liter $\gamma$-catcher
and a 60 cm $\times$ 100 cm stainless steel dark container.  
The prototype was used to do performance tests and background studies,
R\&D for the detector structure and phototube mounting scheme, and to
establish a data analysis effort.
The phototube layout in a simulation of a three-volume
detector is shown in Figure~\ref{fig:reno}.  Each detector would have
a fiducial mass of 15.4 ton using scintillator with density 0.73 gm/cm$^3.$
\par RENO has received support from the South Korean government and good
cooperation from the Y.K. Power Plant Company.  
The expected number of $\neb$'s is about 5000 per day at the near detector
and 100/day at the far detector.  With a systematic error near 1\%,
the project could reach $\sinquq < 0.03$ in 3 years.

\subsection{Angra in Brazil}
The Angra dos Reis reactor complex in Brazil, about 150 km south
of Rio de Janeiro contains two reactors, Angra-I and Angra-II, which
have 2 and 4 GW thermal powers and up times 83\% and 94\% respectively.
The nearby site has high terrain consisting of granite, so both near
and far detectors could have a substantial overburden.  Initial
designs for a $\quq$ experiment involve a near detector, 300 m from
Angra-II,  with 250 m.w.e. overburden, and a far detector, under the peak of
a mountain called ``Morro de Frade", which would provide 2000 m.w.e. at
a distance of 1.5 km.  
Thoughts are to build a 50 ton near detector and a 500 ton far detector,
and concentrate on reducing any bin-to-bin systematic errors.  The
1000 ton KamLAND detector is a proof that large reactor neutrino
detectors are possible.  Unlike KamLAND but like the other new $\quq$
projects, the Angra collaboration plans to build a three-volume detector.
For such a large detector, phototube costs scale as $V^{2/3}$.  A statistical
precision of $\sinquq < 0.006$ could be obtained in three years.

The Angra experiment was originally conceived as a large
$\quq$ detector under a considerable overburden
together with a single reactor in order to obtain an large luminosity but
still have substantial reactor off running.  
A funding request 
to the Brazilian Minister of Science and Technology in 2006 was approved
for initial stages of the project.  
The experiment
will be a long term project which will take advantage of lessons
learned at Double Chooz, Daya Bay and RENO.  In the meantime,
smaller detectors are being constructed with possible applications
toward the monitoring of reactor operations.  The collaboration is
establishing a formal agreement with Eletronuclear for 
permanent access to the site.  They are already authorized to
place one ton of Gd-loaded scintillator provided by LVD 
 near to the reactor for
muon background measurements.  
Other tests have measured noise and singles rate in the vicinity
of the proposed detectors.
\par The next stage is a very near detector with three
volumes of scintillator and a muon to
be placed between 50 and 100 m from the core.  
The current design is a cylinder, 1.3 m high and with a 0.5 m radius
for the target,
1.9 m high and 0.8 m radius for the $\gamma$-catcher, and 3.1 m high
with a 1.4 m radius for the buffer.
\begin{figure}
\vspace*{3pt}
\begin{center}
         \mbox{\epsfig{figure=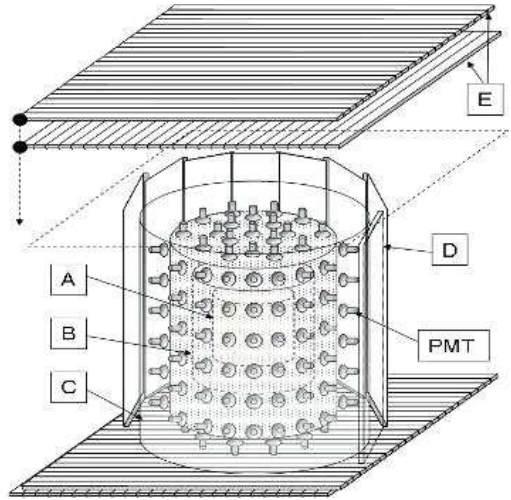,width=8.0cm}}
\caption{Design for the Angra Very Near Detector with
A) target of liquid scintillator and Gd, B) $\gamma$-catcher
of scintillator, C)Buffer of mineral oil and D,E) Muon veto
system of plastic scintillator panels.}
\label{fig:angra}
\end{center}
\end{figure}

\subsection{A comparison of current reactor $\nu$ projects}
Two summaries of some features of the four
current projects are given in Tables \ref{tab:1} and \ref{tab:2}.
These tables were prepared with input from each collaboration's
management in October 2005\cite{bib:pc} and 
may not be up to date.
In any case, the exact size and location
of the detectors is subject to further modifications in design, and
the ``optimistic start dates" need to be taken with a huge clump of
salt.
\begin{table}[ht]
\begin{center}
\caption{Comparison of Detectors for four reactor $\nu$ projects.  
}
\begin{tabular}{|l|l|l|l|l|}\hline
Project & Power (P) & $<P>$ & Location & Detectors \\ \hline
& $GW_{th}$ & $GW_{th}$ &  & km/ton/m.w.e. \\ \hline
& & & & 0.05/1/20 \\
Angra & 6.0 & 5.3 & Brazil & 0.3/50/250 \\
& & & & 1.5/500/2000 \\ \hline
RENO & 17.3 & 16.4 & Korea & 0.15/20/230 \\
& & & & 1.5/20/675 \\ \hline
Daya Bay & 11.6 & 9.9 & China & 0.36/40/260 \\
& (17.4 after 2010) & (14.8 after 2010) & & 0.50/40/260\\
& & & & 1.75/[40x2]/910 \\ \hline
Double & 8.7 & 7.4 & France & 0.27/10.2/90 \\
Chooz & & & & 1.067/10.2/300 \\ \hline
\end{tabular}
\label{tab:1}
\end{center}
\end{table}

\begin{table}[ht]
\begin{center}
\caption{Comparison of Physics for reactor experiments. (* For Daya Bay
after 2010)}
\begin{tabular}{|l|l|r|l|c|c|l|}\hline
Project& Start Date  & GW-t-yr & 90\% CL & for $\dmatmo$ & efficiences & Far event
 \\ 
 & optimistic & (yr) & $\sin^2(2\quq)$ & ($10^{-3}eV^2$) & & rate \\ \hline 
 & & 3900(1) & 0.0070  & & & \\
Angra & 2013(full) & 9000(3) & 0.0060 & 2.5 & 0.8$\times$0.9 & 350,000/yr \\
& & 15000(5) & 0.0055 & 2.5 &  & \\ \hline
RENO & 2009 & 340(1) & 0.03 & 2.0 & 0.8 & 18,000/yr \\ \hline
Daya Bay & 2009 & 3700(3) & 0.008 & 2.5 & 0.75$\times$0.83 & 70,000/yr\\
& & & & & & 110,000/yr$^*$ \\ \hline
Double & 2007(far) & 29(1) & 0.08 & & &  \\
Chooz & 2008(near) & 29(1+1) & 0.04 & 2.5 & 0.8$\times$0.9 & 15,000/yr \\
& & 80(1+3) & 0.025 & & &  \\ \hline
\end{tabular}

\label{tab:2}
\end{center}
\end{table}

\par A comparison of the philosophy of the new reactor projects can
be discerned by a critical examination of Figure \ref{fig:lindner}.  
The thick curve shows the evolution of $\quq$ sensitivity with reactor
luminosity for a particular set of assumptions about the
detector locations, $\dmatmo$ and systematic error.  
That curve showed a transition from near the sensitivity of the rate-only
test to near the sensitivity of the shape-only test between 200 and 2000 
GW-ton-year.
The
Double Chooz and RENO projects aim to quickly improve the limit by
reaching the ``transition" near 200 GW-ton-year.   Daya Bay 
has adopted a goal to work hard to reduce systematic errors 
below the assumptions of Figure \ref{fig:lindner}.  It will reach
$\sinquq \sim 0.01$ with 2000 -ton-year, perhaps by using movable
detectors.  Angra's strategy is to build a much larger far
detector with $ > 10,000$ -ton-year to make it less sensitive
to systematic error.  Depending, of course, on $\dmatmo$, it is
reasonable for the field to have a sensitivity goal of $\sinquq \sim 0.01$,
as might be achievable with the Daya Bay or Angra experiments.
However, as can be seen for Figure \ref{fig:lindner}, the luminosity
requirement for 0.01 is 70 times larger than for 0.03, following the
thick curve.  In that sense, a 0.01 experiment is 70 times harder
than an 0.03 experiment, and the earlier and less expensive Double
Chooz and RENO projects can be valuable steps on the learning curve
for a successful 0.01 experiment.
\par It would be desirable to compare the real schedules of these
four projects.  All four projects have some funding, though not necessarily
enough to reach their design goals (yet).  
Even
though it is the cheapest experiment, Double Chooz' schedule is limited
only by funding.  The other four projects, which will require
considerably more 
civil construction, also have schedules that are probably limited
both by funding and by technical considerations.

\newpage
\section{The near future}
\par The earliest results from reactor experiments may be three years or more
away.  However, at the next Neutrino Telescopes meeting in 2009, observers
will be able to gauge progress by paying attention to the following
subjects:
\begin{itemize}
\item Updated estimates of GW-ton-year as the final detector design
and efficiencies are completed,
\item Liquid scintillator production and stability and attenuation
length studies using large amounts of liquid scintillator,
\item Civil construction issues and, in particular, experience with
costs and schedules,
\item Improved estimates of the background for
cosmogenic sources such as $^9Li$,  (It may be possible to achieve
an improved understanding of the
possible production mechanisms for cosmogenic sources.  In any case,
each experiment should carefully estimate the range of uncertainty of
their background estimates, the impact that uncertainty would have on
the $\quq$ sensitivity, and quantitative methods for measurements that
will lead to a reduction of the uncertainty when data taking is
underway.)
\item Calibration system development and the results of relative
calibration measurements between two or more detectors.
\item Progress in the implementation of movable detectors as
a calibration technique, and evidence as to whether this is a reliable 
method, given the progress or absence of changes when the detectors
are moved.
\end{itemize}
\section{The longer term}
Due to the importance of $\quq$ for CP violation and the mass hierarchy,
a potential long-term program of reactor neutrino measurements 
lies ahead of us.
Results from Double Chooz, Daya Bay, RENO, and later Angra,
will be used to determine
the value of upgrades, additional detectors, and new projects.  An
important factor will be whether the goal becomes further limits on
a small value of $\quq$, or more precise measurements of a non-zero
value.   Statistical precision better than $\delta(\sinquq) < 0.01$
can be imagined, but experience with systematic errors and backgrounds
must be weighed along with the capabilities and needs of accelerator
experiments.  Ideas already exist for more ambitious reactor experiments
to study $\theta_{13}$ further, as well as $\theta_{12}$.  Some examples
are Triple Chooz\cite{bib:triple}, R2D2\cite{bib:r2d2} and 
Hano Hano\cite{bib:learned}.  If such projects become reality, they
will certainly be based on lessons not yet learned by Double Chooz,
Daya Bay, RENO and Angra.

\section{Acknowledgments}
Thanks to the organizers of Neutrino Telescopes 2007 for
the opportunity to discuss reactor neutrinos.  I am indebted to
many colleagues for information in the preparation of this paper,
including Jo{\~a}o dos Anjos, Milind Diwan, Karsten Heeger,
Ernesto Kemp, 
Soo-Bong Kim, Thierry Laserre,
Manfred Lindner, Kam-Biu Luk, 
Guillaume Mention, David Reyna,
Michael Shaevitz and Patricia Vahle.

\end{document}